\documentclass{elsart}

\usepackage{amssymb}
\usepackage{epsfig}

\begin{document}

\begin{frontmatter}

\title{First Investiation of magnetic ground states in the rare-earth intermetallic compounds $R$Al$_{0.9}$Si$_{1.1}$ ($R$ = Ce, Pr, Gd)}

\author[1]{M. H. Jung\corauthref{Jung}},
\author[1]{S. H. Park},
\author[1]{H. C. Kim},
\author[2,3]{Y. S. Kwon}

\corauth[Jung]{Corresponding author. Tel.: +82-42-865-3495;
fax:+82-42-865-3469. E-mail address: mhjung@kbsi.re.kr}

\address[1]{MST, Korea Basic Science Institute, Daejeon 305-333, South Korea}
\address[2]{BK 21 Physics Research Division and Institute of Basic Science, Sungkyunkwan University, Suwon 440-746, South Korea}
\address[3]{Center for Strongly Correlated Material Research, Seoul National University, Seoul 151-742, South Korea}

\begin{abstract}
We report the magnetic properties strongly varying with the
rare-earth elements in the newly found ternary compounds
$R$Al$_{0.9}$Si$_{1.1}$, which crystallize in the tetragonal
$\alpha$-ThSi$_2$-type structure. For $R$ = Ce the alloy has a
weak ferromagnetism below 11 K and for $R$ = Pr it orders
ferromagnetically at 17 K, while for $R$ = Gd it is
antiferromagnetic with $T_{\rm N}$ = 30.5 K. In addition, we find
no field effect on $T_{\rm N}$ of $R$ = Gd because of the large
internal mean field, but significant changes in the magnetic
properties of $R$ = Ce and Pr.
\end{abstract}

\end{frontmatter}

There have been studies of Ce-based ternary compounds in the
Ce-Al-(Si,Ge) phase diagram \cite{1}. It was found for
CeAl$_x$(Si,Ge)$_{2-x}$ that the crystal structure changes as the
Al/Si(Ge) ratio varies \cite{2}. The alloys of $x$ = 1 and 1.2
have the tetragonal $\alpha$-ThSi$_2$-type structure, while the
alloy of $x$ = 1.5 has the hexagonal AlB$_2$-type structure. The
magnetic properties were reported to depend on the crystal
chemistry \cite{3}; CeAlSi is ferromagnetic with a Curie
temperature of 7.1 K, whereas CeAlGe orders antiferromagnetically
at 4 K. In the present work, we have extended this investigation
to other ternary compounds as changing the rare-earth elements,
which are $R$Al$_{0.9}$Si$_{1.1}$ ($R$ = Ce, Pr, and Gd) and find
that the magnetic properties are strongly varying with the
rare-earth elements.

The single crystals were synthesized by high-temperature flux
method. Electron-probe microanalysis showed the composition of
$R$Al$_{0.9}$Si$_{1.1}$ within an error of $\pm 0.05$ for $R$ = Gd
and $\pm 0.03$ for $R$ = Ce and Pr without any impurity phase.
X-ray powder diffraction pattern revealed that the samples are
single phased with the tetragonal $\alpha$-ThSi$_2$-type
structure. The magnetic susceptibility was measured in a field of
100 G using a Quantum Design superconducting quantum interference
device. The specific heat was taken with a Quantum Design physical
property measurement system.

\begin{figure}
\begin{center}
\includegraphics[width=0.7\linewidth]{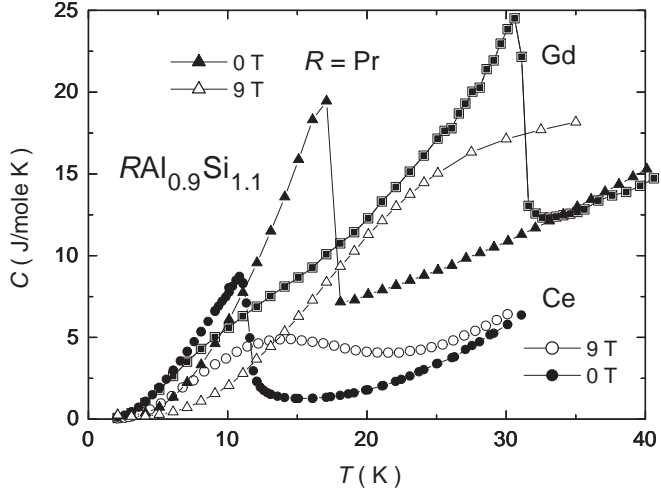}
\caption{Specific heat $C(T)$ vs. temperature for $R$ = Ce, Pr,
and Gd in $R$Al$_{0.9}$Si$_{1.1}$. The solid symbols are
zero-field data and the open symbols are 9T data.} \label{fig2}
\end{center}
\end{figure}

The inverse magnetic susceptibility $H/M$ vs. temperature is shown
in Fig. 1. A linear behavior of the Curie-Weiss law is observed
above 100 K and a deviation from the linear behavior occurs below
50 K. The effective magnetic moments are estimated to be $\mu
_{eff}$ = 2.52(5) $\mu_B$, 3.31(8) $\mu_B$, and 8.10(8) $\mu_B$
for $R$ = Ce, Pr, and Gd, respectively. These values show that the
rare-earth ions are in the normal trivalent state. The
paramagnetic Curie temperatures are $\theta_P$ = 6.91(0) K,
26.37(4) K, and $-$142.43(3) K for $R$ = Ce, Pr, and Gd,
respectively, possibly indicating the development of different
types of magnetic exchange interaction. The small value of
$\theta_P$ for CeAl$_{0.9}$Si$_{1.1}$ might be related with a weak
ferromagnetism, while the negative value of $\theta_P$ for
GdAl$_{0.9}$Si$_{1.1}$ might be responsible for an
antiferromagnetic order.

These different types of magnetic exchange could be achieved at
low temperatures. In the inset of Fig. 1, we observe a peak in the
zero-field cooled (ZFC) curve and a cusp in the field cooled (FC)
curve at $T_{\rm wF}$ = 11 K, which further confirms weak
ferromagnetism in CeAl$_{0.9}$Si$_{1.1}$. The magnetic behavior of
PrAl$_{0.9}$Si$_{1.1}$ could be also understood in a way similar
to that of CeAl$_{0.9}$Si$_{1.1}$, because there is a difference
between ZFC and FC susceptibilities below $T_{\rm C}$ = 17 K.
However, we observe a rapid saturation of magnetization to a value
of almost full moment $\sim 3 \mu_B$ at 0.4 T (not shown here),
indicating a ferromagnetic ordering. It is worthwhile to mention
here that this difference could be associated with a spin-glass
transition. On the other hand, GdAl$_{0.9}$Si$_{1.1}$ has no
difference between ZFC and FC susceptibilities and orders
antiferromagnetically below $T_{\rm N}$ = 30.5 K. The
magnetization increases linearly with magnetic field (not shown
here).

\begin{figure}
\begin{center}
\includegraphics[width=0.7\linewidth]{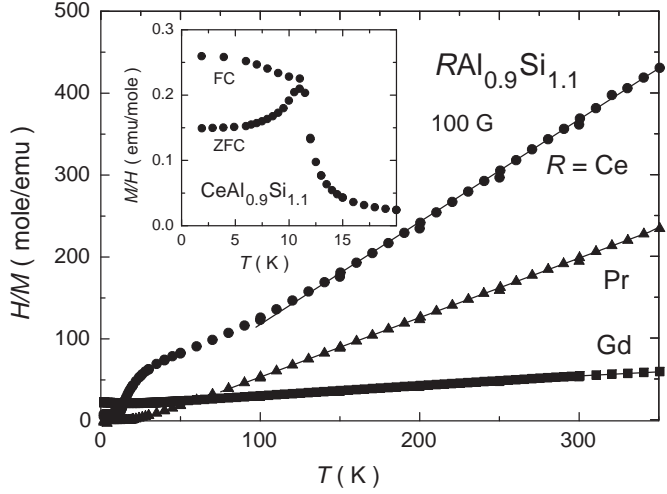}
\caption{Inverse magnetic susceptibility $H/M$ vs. temperature for
$R$ = Ce, Pr, and Gd in $R$Al$_{0.9}$Si$_{1.1}$. The solid lines
represent linear fits of the Curie-Weiss law. The inset shows the
low-temperature susceptibility $M/H$ of $R$ = Ce.} \label{fig1}
\end{center}
\end{figure}

Figure 2 represents the specific heat $C(T)$ vs. temperature. A
$\lambda$-type anomaly is observed in all the compounds
$R$Al$_{0.9}$Si$_{1.1}$ at the magnetic transition temperatures:
$T_{\rm wF}$ = 11 K, $T_{\rm C}$ = 17 K, and $T_{\rm N}$ = 30.5 K
for $R$ = Ce, Pr, and Gd, respectively. As the magnetic is
applied, the magnetic transition temperatures for $R$ = Ce and Pr
are increased, whereas the transition temperature for $R$ = Gd is
unchanged. The former is characteristic of ferromagnetic
materials. In the latter case, one should note that Gd$^{3+}$ ions
(L = 0 and J = S) in GdAl$_{0.9}$Si$_{1.1}$ have a large internal
mean field and thus an effect of external field could be
negligible in total magnetic exchange. We may depress $T_{\rm N}$
in a very high magnetic field beyond our present measurement
range.

This work was supported by the Korea Science and Engineering
Foundation through the Center for Strongly Correlated Materials
Research at SNU and the NRL project of the Korea Ministry of
Science and Technology.

\end{document}